\documentclass[aps,prb,superscriptaddress,twocolumn]{revtex4-2}
\usepackage{graphicx,color}
\usepackage{verbatim}
\usepackage{amssymb}   % for math
\usepackage{amsmath}
\usepackage{amsfonts}
\usepackage{mathdots}
\usepackage{hyperref}
\usepackage{multirow}
\usepackage{epsfig}
\usepackage{hyperref}
\usepackage{xcolor}
\usepackage{cancel}
\usepackage{booktabs}
\usepackage{array}
\usepackage{array, makecell} 
\definecolor{myblue}{RGB}{46, 48,146}
\usepackage{multirow}
\hypersetup{colorlinks=true,linkcolor=myblue,citecolor=myblue,urlcolor=myblue, linktocpage}
\usepackage{braket}
\usepackage{bm}
\usepackage{multirow}

\begin{document}
	\title{Probing Loop Currents and Collective Modes of Charge Density Waves in Kagome Materials with NV Centers}

	\author{Ying-Ming Xie}\thanks{yingming.xie@riken.jp}
	\affiliation{RIKEN Center for Emergent Matter Science (CEMS), Wako, Saitama 351-0198, Japan} 	
	\author{Naoto Nagaosa} \thanks{nagaosa@riken.jp}
	\affiliation{RIKEN Center for Emergent Matter Science (CEMS), Wako, Saitama 351-0198, Japan} 
	\affiliation{Fundamental Quantum Science Program, TRIP Headquarters, RIKEN, Wako 351-0198, Japan} 	
	
	\date{\today}
	\begin{abstract}
Recently, the unconventional charge density wave (CDW) order with loop currents has attracted considerable attention in the Kagome material family AV$_3$Sb$_5$ (A = K, Rb, Cs). However, experimental signatures of loop current order remain elusive. In this work, based on the mean-field free energy, we analyze the collective modes of unconventional CDW order in a Kagome lattice model. Furthermore, we point out that phase modes in the imaginary CDW (iCDW) order with loop current orders result in time-dependent stray fields. We thus propose using nitrogen-vacancy (NV) centers to detect these time-dependent stray fields, providing a potential experimental approach to identifying loop current order.

	\end{abstract}
	
	\maketitle
	In condensed matter physics, electron correlations can lead to the emergence of new states of matter with various electronic and spin orders. Loop current order is an intriguing electronic order that leads to local current loops in a material, spontaneously breaking time-reversal symmetry.  In the late 1990s, in an effort to understand the pseudogap phase, C. M. Varma proposed that loop current order could exist within the CuO$_2$ unit cell of cuprates \cite{Varma1997, Varma1999}. However, the experimental identification of loop current order in cuprates remains challenging and controversial \cite{Bourges, Croft2017, Gheidi2020}.

Recently, the newly discovered Kagome materials AV$_{\text{3}}$Sb$_{\text{5}}$ (A = K, Rb, and Cs) have emerged as an alternative platform for exploring loop current order \cite{Neupert2022, Wilson2024, Kun2025}. In the AV$_{\text{3}}$Sb$_{\text{5}}$ family, various experimental signatures suggest the presence of unconventional charge density waves (CDWs) with time-reversal symmetry breaking, as indicated by scanning tunneling microscopy \cite{Jiang2021}, spin resonance ($\mu$SR) \cite{Mielke2022}, magneto-optical Kerr effect \cite{Xu2022}, and other techniques \cite{Neupert2022, Wilson2024, Kun2025}.  
To break time-reversal symmetry, theoretically proposed CDW orders in these Kagome materials often involve loop current order, where the order parameter is imaginary \cite{Titus2021, Leon2021, FENG2021, Lin2021, Baek2024}, leading to so-called imaginary charge density waves (iCDW). Despite significant progress in this field, there is still no definitive experimental evidence for loop current order in AV$_{\text{3}}$Sb$_{\text{5}}$. Moreover, recent high-resolution polar Kerr measurements have instead suggested the absence of time-reversal symmetry breaking in the charge-ordered state of CsV$_3$Sb$_5$ \cite{Saykin131, Farhang2023}.  The study of loop current order in Kagome materials continues to attract growing interest \cite{Rina2024, Ruiqing2024, fernandes2025}.

On the other hand, understanding the collective excitations of an ordered state is crucial for gaining deeper insights into its nature. Loop current fluctuations are known to give rise to various intriguing physical phenomena in strongly correlated systems \cite{Varma2011, Zheyang2023, Grgur2024}. Recently, the amplitude modes of charge order in CsV$_3$Sb$_5$ have been experimentally investigated \cite{Liu2022, Doron2023}, and the ultrafast control of charge order in Kagome metals has been numerically studied \cite{Yuping2024}.  
Although many previous theoretical works have modeled iCDW order with loop currents \cite{Titus2021, Leon2021, FENG2021, Lin2021, Baek2024}, a simple theoretical analysis of their collective excitations remains elusive. Key open questions include whether the collective excitations of iCDW exhibit novel properties and whether these excitations can serve as a probe for detecting loop current order. These questions, along with recent experimental progress on Kagome materials, motivate the present study.  Additionally, we are inspired by our previous work on real triple-Q CDW order parameters, where we analyzed phase shifts and band geometry effects \cite{Naoto2024}.  We now are curious about any interesting properties from the imaginary triple-Q CDW order related to the phase degree of freedom.

In this work, we theoretically analyze the collective excitations of triple-Q order parameters, including both real CDW (rCDW) and iCDW order, in a Kagome lattice model. We explicitly obtain the phase and amplitude modes and find a crucial distinction: in iCDW order, phase and amplitude modes mix, whereas in rCDW order, they remain decoupled.  Furthermore, we propose that phase mode excitations can serve as a probe for detecting loop currents in the iCDW phase using nitrogen-vacancy (NV) centers, as loop current fluctuations generate magnetic noise. The proposed experimental setup is illustrated in Fig.~\ref{fig:fig1}.  In the past, NV centers have been used or proposed for probing various correlated orders, including currents and chirality fluctuation in high temperature superconductors \cite{Lee1991}, antiferromagnetic order \cite{Appel2019, Flebus2018, Finco2021}, conventional superconducting efffects \cite{Ron2018, Demler2022, Monge2023}, and quantum spin liquids \cite{Inti2022, Lee2023} (see Refs.~\cite{Casola2018, Rovny2024} for a review). Our proposal provides new motivation for applying NV center detection in the search for loop current order.

\begin{figure}
    \centering
    \includegraphics[width=1\linewidth]{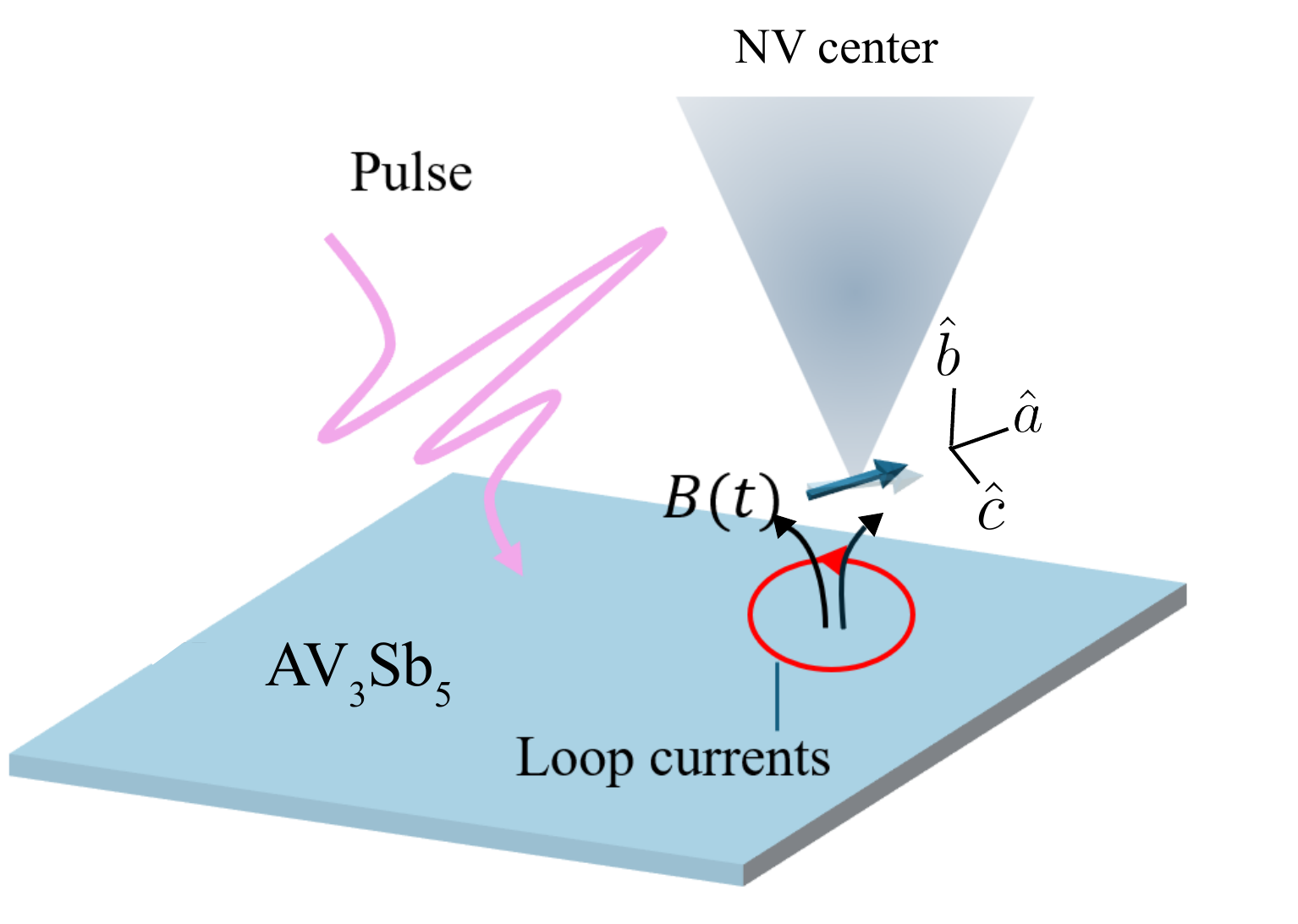}
    \caption{Schematic of the proposed setup to detect the loop current order using NV centers. In this setup, time-dependent magnetic fields arise from loop current fluctuations induced by a laser pulse. The corresponding magnetic noise can be detected by the NV center.
   }
    \label{fig:fig1}
\end{figure}

\noindent{\bf Results}

\noindent {\bf Kagome lattice model with triple-Q CDW.} To set the stage, we briefly recall the Kagome lattice model. The Kagome lattice is shown in Fig.~\ref{fig:fig2}(a), with the corresponding point group symmetry being $D_{6h}$. Each unit cell consists of three sublattices: A, B, and C. The unit vectors are given by $\bm{a}_1 = (2,0)a$ and $\bm{a}_2 = (1,\sqrt{3})a$, where $a$ is the bond length. The energy bands with nearest-neighbor hopping $t$ are shown in Fig.~\ref{fig:fig2}(b). In the AV$_3$Sb$_5$ family, the chemical potential is near the Van Hove singularity points (red dashed line), which plays a crucial role in inducing various symmetry-breaking orders.  Fig.~\ref{fig:fig2}(c) depicts the Brillouin zone. The Van Hove singularity points are located at three distinct $M$ points: $M_1$, $M_2$, and $M_3$. The Bloch wavefunctions at the $M_1$, $M_2$, and $M_3$ points arise from the A, B, and C sublattices, respectively.

The coupling between the Van Hove singularity points leads to charge instability. In experiments on AV$_3$Sb$_5$ materials, a $2 \times 2$ CDW order has been observed using scanning tunneling microscopy (STM) \cite{Jiang2021, Zhao2021, Chen2021}. As demonstrated previously \cite{FENG2021, Rina2024}, a natural way to generate such a CDW order is through a triple-Q coupling, with $\bm{Q}_1 = \left(\frac{\sqrt{3}}{2}, -\frac{1}{2}\right)|M|$, $\bm{Q}_2 = \left(-\frac{\sqrt{3}}{2}, -\frac{1}{2}\right)|M|$, and $\bm{Q}_3 = (0, 1)|M|$, where $|M| = \frac{\pi}{\sqrt{3}a}$ [Fig.~\ref{fig:fig2}(c)]. The three $\bm{Q}$ vectors induce coupling between the $M$ points, and the folded Brillouin zone is highlighted by the gray lines in Fig.~\ref{fig:fig2}(c).  The ansatz for the CDW order is given by:
\begin{eqnarray}
	\Delta_{CDW}(\bm{r}) = \begin{pmatrix}
		0 & \Delta_{\bm{Q}_3}(\bm{r}) & \Delta^*_{\bm{Q}_2}(\bm{r}) \\
		\Delta_{\bm{Q}_3}(\bm{r})^* & 0 & \Delta_{\bm{Q}_1}(\bm{r}) \\
		\Delta_{\bm{Q}_2}(\bm{r}) & \Delta^*_{\bm{Q}_1}(\bm{r}) & 0
	\end{pmatrix}, \nonumber \\
\end{eqnarray}
where $\Delta_{\bm{Q}_j}(\bm{r}) = |\Delta_{\bm{Q}_j}|e^{i(\bm{Q}_j \cdot \bm{r} + \theta_j)}$, and the basis is $(c_{A}, c_{B}, c_{C})$, with $c$ being the electron annihilation operator. Here, we define the phase degree of freedom of order parameter $\Delta_{\bm{Q}_j}$ as $\theta_{j}$.

It is important to note that the CDW order parameters can be complex due to the phase degree of freedom $\theta_{j}$. Because of the commensurability of this triple-Q CDW, the values of $\theta_{j}$ are not arbitrary and must be determined by minimizing the free energy, as discussed in the following section.

%The resulting patch Hamiltonian near $M$ points is
%\begin{equation}
%H(\bm{k})=\begin{pmatrix}
%\epsilon_{1}(\bm{k})&|\Delta_{\bm{Q}_3}|e^{i\theta_3}& |\Delta_{\bm{Q}_2}|e^{-i\theta_2}\\
%|\Delta_{\bm{Q}_3}|e^{-i\theta_3}&\epsilon_{2}%(\bm{k})& |\Delta_{\bm{Q}_1}|e^{i\theta_1}\\
%|\Delta_{\bm{Q}_2}|e^{i\theta_2}&|\Delta_{\bm{Q}_1}|e^{-i\theta_1}&\epsilon_3(\bm{k})
%\end{pmatrix}.
%\end{equation}
%where $\epsilon_1(\bm{k})=-\mu-2\sqrt{3} t k_xk_y-2tk_x^2$, $\epsilon_2(\bm{k})=-\mu-3 t k_y^2+2tk_x^2$, $\epsilon_3(\bm{k})=-\mu+2\sqrt{3} t k_xk_y-2tk_x^2$.  The momentum $\bm{k}$ is measured with respect to the $M_{\alpha}$ points (note that the three $M$ points would be folded as the same point in the foled Brillouin zone). 
%The $H(\bm{k})$ severves a simple low-energy Hamiltnoian for the aforementioned triple-Q CDW states in this Kagome lattice.

\begin{figure}
    \centering
    \includegraphics[width=0.8\linewidth]{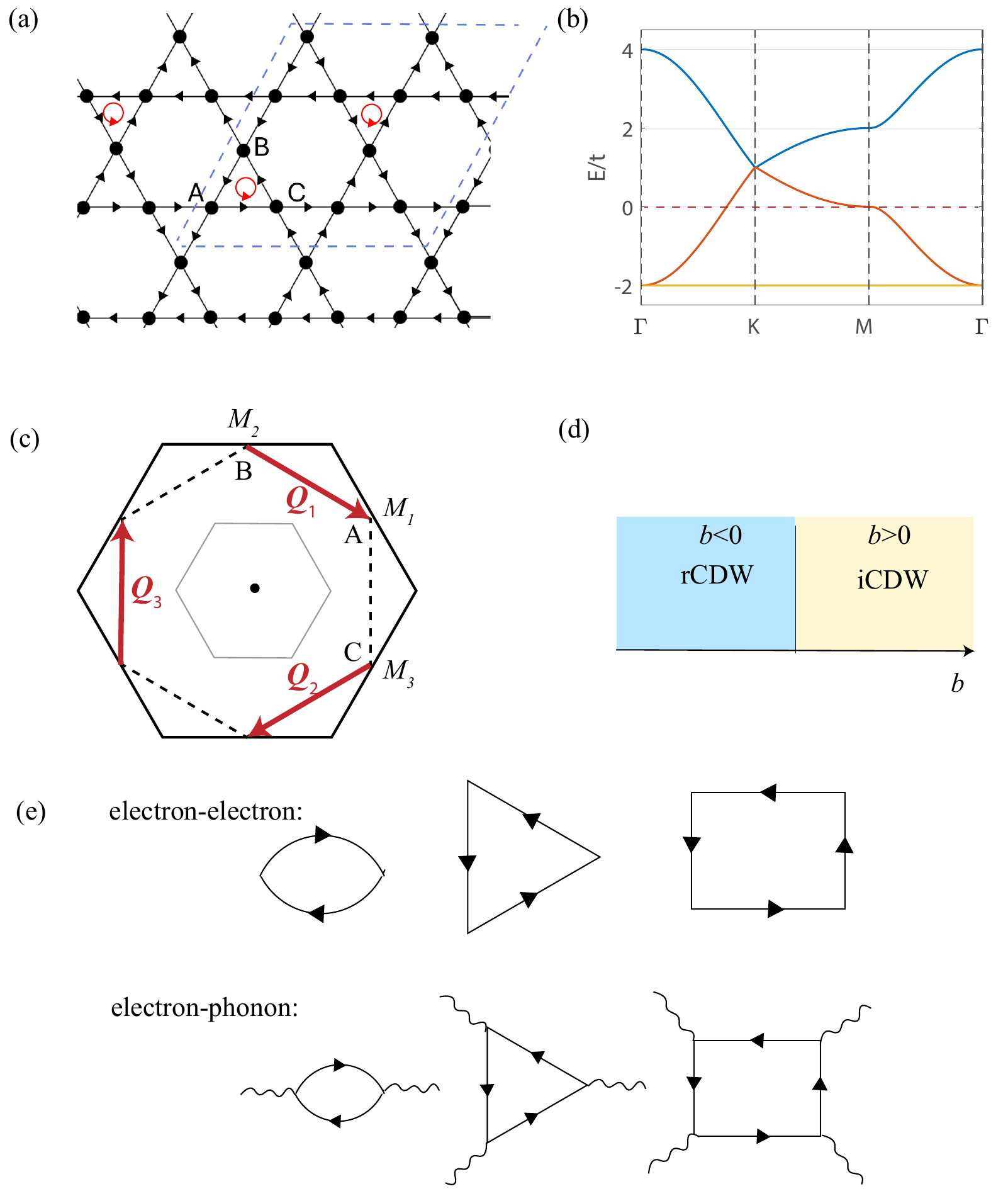}
    \caption{ Kagome lattice model. (a) The Kagome lattice with triple-Q CDW order. Here, A, B, and C label the three sublattices, and the phases are indicated by the arrows in the iCDW phase. The red circles are the loop currents in the small triangle  plaquettes. (b) The single-particle electronic band structure of the Kagome lattice toy model. (c) The schematic plot of the original Brillouin zone (black solid line) and the folded Brillouin zone (gray line). The $M$ points, with wavefunctions localized on the A, B, and C sublattices, are highlighted. (d) A schematic phase diagram controlled by $b$ in Eq.~\ref{free}. When $b<0$ ($b>0$), the rCDW (iCDW) phase tends to be favored. (e) The Feynman diagram representation for the free energy terms (top panel for electron-electron interaction, bottom panel for electron-phonon interaction).
 }
    \label{fig:fig2}
\end{figure}

 Phenomenologically, the mean-field free energy can be expanded as a series in powers of the order parameter $\Delta_{\bm{Q}}$. In the triple-Q case, the allowed terms are of the form $\Delta^{n_1}_{\bm{Q}_1}\Delta^{n_2}_{\bm{Q}_2}\Delta^{n_3}_{\bm{Q}_3}$, where $\sum_{n_i} n_i \bm{Q}_j = \bm{G}$. Here, $n_i$ are integers, $\bm{G}$ represents the reciprocal lattice vectors, and we define $\Delta^{n_1}_{\bm{Q}_i} = (\Delta_{\bm{Q}_i})^{n_i}$ when $n_i \geq 0$, and $\Delta^{n_1}_{\bm{Q}_i} = (\Delta^*_{\bm{Q}_i})^{|n_i|}$ when $n_i < 0$. For the CDW with triple-Q vectors shown in Fig.~\ref{fig:fig2}, the free energy up to the fourth order is given by \cite{McMillan1975,  Nagaosa1984, vanWezel_2011, Titus2021, Leon2021}
\begin{eqnarray}
    F &\approx & \sum_{\alpha} \left( b \cos 2\theta_{\alpha} - \lambda_1' \right) |\Delta_{\bm{Q}_{\alpha}}|^2 \nonumber \\
    && + \lambda_2 |\Delta_{\bm{Q}_1}| |\Delta_{\bm{Q}_2}| |\Delta_{\bm{Q}_3}| \cos(\theta_1 + \theta_2 + \theta_3) \nonumber \\
    &&+ \lambda_{3} |\Delta_{\bm{Q}_1}| |\Delta_{\bm{Q}_2}| |\Delta_{\bm{Q}_3}| \cos(\theta_1) \cos(\theta_2)\cos(\theta_3)+\nonumber\\
    && + u_1 \left( \sum_{\alpha} |\Delta_{\bm{Q}_\alpha}|^4 \right)+ u_2 \sum_{\alpha \neq \beta} |\Delta_{\bm{Q}_\alpha}|^2 |\Delta_{\bm{Q}_\beta}|^2. \label{free}
\end{eqnarray}
According to ref.~\cite{Leon2021},   the phenomenological free energy can also be derived from the specific electron-electron interactions (see Supplementary Material for details), where $\lambda_3$  is zero. To reduce number of parameters,  we adopt the convention of ref.~\cite{Leon2021} with $\lambda_3=0$,  and the $C_3$ symmetry  that requires $\theta_1 = \theta_2 = \theta_3 = \theta_0$ and $|\Delta_{\bm{Q}_1}| = |\Delta_{\bm{Q}_2}| = |\Delta_{\bm{Q}_3}|$ [Supplementary Material Sec. I].  Since the dominant term is the second-order term, the free energy is minimized when $\theta_0 = \frac{\pi}{2}$ or $-\frac{\pi}{2}$ for $b > 0$, which corresponds to the iCDW phase. On the other hand, the free energy is minimized when $\theta_0 = 0$ or $\pi$ for $b < 0$, which corresponds to the rCDW phase.  The phase diagram is sketched in Fig.~\ref{fig:fig2}(d). The sign of $b$ is closely related to the interactions. Note that if $\lambda_2 < 0$ ($\lambda_2 > 0$), the rCDW phase with $\theta_0 = 0$ ($\theta_0 = \pi$) is more favorable than with $\theta_0 = \pi$ ($\theta_0 = 0$ ) due to the third-order term.

We emphasize that the microscopic origin of the interactions driving the CDW is not the focus of this work. The phenomenological free energy in Eq.~\eqref{free} applies to both electron-electron and electron-phonon interactions. Depending on the type of interaction (electron-electron or electron-phonon), the second- to fourth-order terms in the free energy $F$ can be represented by the Feynman diagrams shown in Fig.~\ref{fig:fig2}(e).

\vspace{1\baselineskip}
\noindent{\bf Collective Modes Analysis.} We are now ready to analyze the collective modes based on the phenomenological free energy. By incorporating the fluctuations of the CDW order parameters, the Lagrangian is given by
\begin{eqnarray}
\mathcal{L} &&= \sum_{\alpha} \kappa_0|\partial_{\tau}\Delta_{\bm{Q}_{\alpha}}|^2 + \kappa_1|\nabla \Delta_{\bm{Q}_{\alpha}}|^2 + (b\cos 2\theta_{\alpha} - \lambda_1')|\Delta_{\bm{Q}_{\alpha}}|^2 \nonumber\\
&&+ \lambda_2|\Delta_{\bm{Q}_1}| |\Delta_{\bm{Q}_2}| |\Delta_{\bm{Q}_3}| \cos(\theta_1 + \theta_2 + \theta_3) + \nonumber\\
&& u_1 \sum_{\alpha} |\Delta_{\bm{Q}_\alpha}|^4 + u_2 \sum_{\alpha \neq \beta} |\Delta_{\bm{Q}_\alpha}|^2 |\Delta_{\bm{Q}_\beta}|^2.
\end{eqnarray}
The first two terms in $\mathcal{L}$ describe the time and spatial fluctuations, respectively. To obtain the phase and amplitude modes, we rewrite the order parameters as $\Delta_{\bm{Q}_{\alpha}, \bm{q}} \approx \Delta_{\bm{Q}_\alpha}(1 + \mathcal{A}_{\alpha}(\bm{q})) e^{i(\theta_0+\theta_{\alpha}(\bm{q}))}$. Using the $C_3$ symmetry, the amplitude modes can be projected into the $A$ and $E_{1,2}$ irreducible representations:
\begin{eqnarray}
\mathcal{A}^{(A)}_{q} &=& \frac{1}{\sqrt{3}} (\mathcal{A}_1(q) + \mathcal{A}_2(q) + \mathcal{A}_3(q)), \\
\mathcal{A}^{(E_1)}_q &=& \frac{1}{\sqrt{2}} (\mathcal{A}_1(q) - \mathcal{A}_3(q)), \\
\mathcal{A}^{(E_2)}_q &=& \frac{1}{\sqrt{6}} (\mathcal{A}_1(q) - 2\mathcal{A}_2(q) + \mathcal{A}_3(q)),
\end{eqnarray}
where $q = (\bm{q}, \omega)$ and $\omega$ is the frequency. The phase modes can be projected into the following channels:
\begin{eqnarray}
\theta^{(A)}_{q} &=& \frac{1}{\sqrt{3}} (\mathcal{\theta}_1(q) + \mathcal{\theta}_2(q) + \mathcal{\theta}_3(q)), \\
\theta^{(E_1)}_q &=& \frac{1}{\sqrt{2}} (\theta_{1}(q) - \theta_3(q)), \\
\theta^{(E_2)}_q &=& \frac{1}{\sqrt{6}} (\theta_{1}(q) - 2\theta_2(q) + \theta_3(q)).
\end{eqnarray}

As shown in Supplementary Material Sec.~II, the fluctuation part of the Lagrangian can be simplified as
\begin{widetext}
\begin{eqnarray}
\mathcal{L}_{fluc}(\omega, \bm{q}) &=& (\kappa_1 \bm{q}^2 - \kappa_0 \omega^2 + 4(u_1 + 2u_2) |\Delta_Q|^2) \mathcal{A}^{(A)}_{q} \mathcal{A}^{(A)}_{-q} + (\kappa_1 \bm{q}^2 - \kappa_0 \omega^2 + 2|b| - \frac{3}{2} \lambda_2 |\Delta_Q| \cos(3\theta_0)) \theta_q^{(A)} \theta_{-q}^{(A)} \nonumber \\
&& + \frac{3}{2} \lambda_2 |\Delta_Q| \sin(3\theta_0) (\mathcal{A}^{(A)}_q \theta^{(A)}_{-q} + \mathcal{A}^{(A)}_{-q} \theta^{(A)}_q) + (\kappa_1 \bm{q}^2 - \kappa_0 \omega^2 + 2|b|) (\theta_q^{(E_1)} \theta_{-q}^{(E_1)} + \theta_q^{(E_2)} \theta_{-q}^{(E_2)}) \nonumber \\
&& + (\kappa_1 \bm{q}^2 - \kappa_0 \omega^2 - \frac{3}{2} \lambda_2 \cos(3\theta_0) |\Delta_Q| + 4(u_1 - u_2) |\Delta_Q|^2) (\mathcal{A}^{(E_1)}_q \mathcal{A}^{(E_1)}_{-q} + \mathcal{A}^{(E_2)}_q \mathcal{A}^{(E_2)}_{-q}).
\end{eqnarray}
\end{widetext}
Here, the $A$- and $E$-modes are separated due to the $C_3$ symmetry. As expected, both the phase and amplitude modes are gapped in this commensurate triple-Q CDW. More intuitively, the free energy landscape exhibits local minima in the complex $\Delta$ plane, and fluctuations around these local minima give rise to the massive modes (see Fig.~\ref{fig:fig3}(a) for an illustration). Interestingly, we observe a mixing term between the phase mode and amplitude mode, $\propto \sin(3\theta_0)(\mathcal{A}^{(A)}_q \theta^{(A)}_{-q} + \mathcal{A}^{(A)}_{-q} \theta^{(A)}_q)$. This term is finite in the iCDW phase with $\theta_0 = \frac{\pi}{2}, -\frac{\pi}{2}$, and vanishes for rCDW with $\theta_0 = 0, \pi$. For rCDW, a direct mixing term between phase and amplitude modes that breaks time-reversal symmetry is not allowed. An example of such a forbidden term is  
$(\beta_1 + \beta_2 \cos(3\theta_0))(\mathcal{A}^{(A)}_q \theta^{(A)}_{-q} + \mathcal{A}^{(A)}_{-q} \theta^{(A)}_q)$. Note that under the time-reversal operation, $\theta_{q} \mapsto -\theta_{-q}$, $\mathcal{A}_{q} \mapsto \mathcal{A}_{-q}$, and $\theta_0 \mapsto -\theta_0$. It can also be seen that all terms in $\mathcal{L}_{\text{fluc}}(\omega, \bm{q})$ are time-reversal even.

 The mixed Higgs and phase modes in the iCDW (at $\bm{q}=0$, i.e., $M$ points) respect
\begin{equation}
\begin{vmatrix}
-\kappa_0 (\omega^{(A)})^2 + 4(u_1 + 2u_2) |\Delta_Q|^2 & \frac{3}{2} \lambda_2 \sin(3\theta_0) |\Delta_Q| \\
\frac{3}{2} \sin(3\theta_0) \lambda_2 |\Delta_Q| & -\kappa_0 (\omega^{(A)})^2 + 2|b|
\end{vmatrix} = 0.
\end{equation}
This gives
\begin{eqnarray}
\kappa_0 (\omega_{\pm}^{(A)})^2 &=& |b| + 2(u_1 + 2u_2) |\Delta_Q|^2 \nonumber \\
&& \pm \sqrt{(|b| - 2(u_1 + 2u_2) |\Delta_Q|^2)^2 + \frac{9}{4} \lambda_2^2 |\Delta_Q|^2}. \nonumber \\
\end{eqnarray}
Here, $\omega_{\pm}^{(A)}$ denotes the energy of the mixed phase-amplitude mode. We summarize the expected collective mode spectrum for rCDW and iCDW in Fig.~\ref{fig:fig3}(b). Note that the ordering of these modes along the frequency axis, depending on parameters,  may differ from that shown in Fig.~\ref{fig:fig3}(b).

\begin{figure}
    \centering
    \includegraphics[width=1\linewidth]{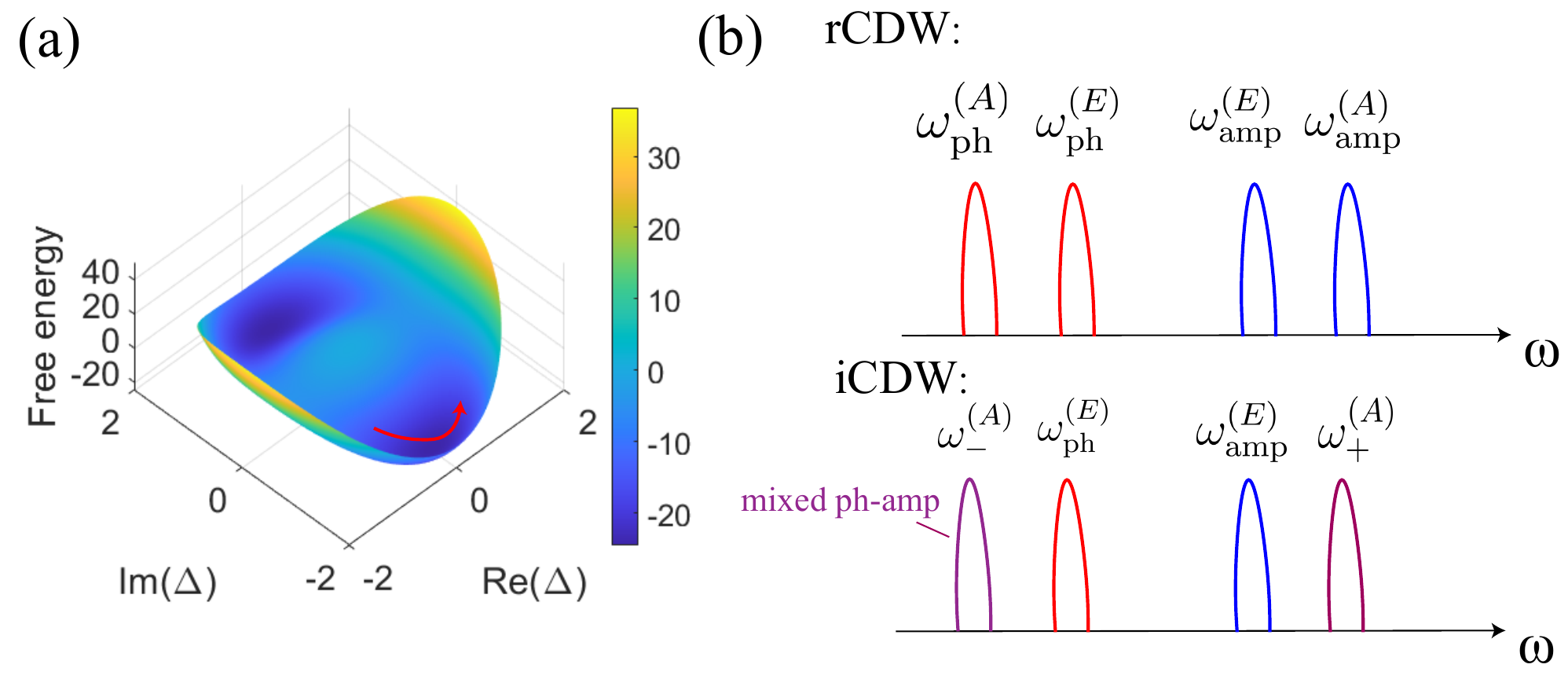}
    \caption{The CDW collective excitations. (a) The free energy landscape from Eq.~\eqref{free} with parameters $b=1$, $\lambda_1'=5$, $\lambda_2=0.1$, $\lambda_3=0$, $u_1=0.5$, and $u_2=0.5$. The red arrow represents the fluctuation of the complex order parameter. (b) A schematic plot of the spectral arrangement of phase (red) and amplitude (blue) modes for rCDW and iCDW orders. Here, the phase and amplitude modes in the $A$ channel mix (purple),  while they remain uncoupled for $E$ modes.
  }
    \label{fig:fig3}
\end{figure}

\vspace{1\baselineskip}
\noindent{\bf Loop current detection with NV Centers.}  
The phase fluctuation in the iCDW phase is particularly interesting because it implies that the magnetic flux associated with the loop current order is dynamic, generating a time-varying magnetic stray field. The iCDW order effectively mediates imaginary inter-sublattice hopping, which gives rise to current within each bond \cite{Titus2021}. The phase of iCDW order or bound current direction  is illustrated with black arrows in Fig.~\ref{fig:fig2}(a) \cite{FENG2021, Rina2024}. The loop currents appear when the arrows are connected in a clockwise or counterclockwise direction within each plaquette. As shown in previous works \cite{FENG2021, Rina2024}, the net loop current for the iCDW arises from small triangular plaquettes with counterclockwise arrows. According to the Peierls substitution principle, the flux within these small triangular plaquettes is given by the sum of the phases:
$\Phi \propto \sum_{\alpha} \theta_{\alpha} = \sqrt{3} \theta^{(A)}$. Note that such flux is absent for rCDW, as the order paramter is real and does not break time-reversal symmetry. 
When the $A_1$ phase mode is excited, we expect that the average flux of the system with iCDW order will oscillate in time as
\begin{equation}
\tilde{\Phi}(t) = \tilde{\Phi}_0 + \delta \tilde{\Phi} \sin(\omega^{(A)}_{\text{ph}} t).
\end{equation}
Here, $\tilde{\Phi}_0$ is the static flux, and $\delta \tilde{\Phi}$ represents the amplitude of the fluctuating part.  This dynamic flux is expected to generate magnetic noise, which can potentially be detected by an NV center detector \cite{Rovny2024}.

The coupling between the stray field $\bm{B}(t)$ and the NV center is described by the Hamiltonian
$H_{NV} = D_{\text{zfs}} \hat{S}_{\hat{a}}^2 - \gamma_e B_0 \hat{S}_{\hat{a}} - \gamma_e \bm{B}(t) \cdot \bm{S} $) \cite{Rovny2024}.
Here, $\hat{a}$ is along the quantization axis of the NV center, $D_{\text{zfs}} = 2\pi \times 2.87$ GHz is the zero-field splitting, $\gamma_e = -2\pi \times 28.02$ GHz·T$^{-1}$ is the electron g-factor, and $B_0$ is a static magnetic field. The NV center exhibits two operational modes: $T_1$ spectroscopy and $T_2$ spectroscopy. Specifically, the $T_1$ relaxation time describes how quickly the spin population of the NV center in the $m_s = \pm 1$ states decays to the lower-energy $m_s = 0$ state, while $T_2$ describes the coherent dephasing time of the NV spin. The $T_2$ spectroscopy is primarily sensitive to low-frequency noise (Hz - MHz), while $T_1$ spectroscopy is sensitive to fluctuating magnetic fields at the NV's Larmor frequency (up to GHz or sub-THz regions). For example, recently proposed magnon noise (in the range of 10 to 100 GHz) \cite{Flebus2018} was observed experimentally using NV relaxometry \cite{Finco2021}. Moreover, recent advances in high-field NV spectroscopy have enabled the detection of magnetic noise in the sub-THz region \cite{Fortman, Sandor2024}. In our scenario, if the iCDW is commensurate, the phase mode gap is determined by the strength of the lattice pinning of the CDW, typically in the sub-THz region. In fact, it has been demonstrated that the CDW can be tuned into an incommensurate state by doping Kagome materials such as CsV$_3$Sb$_{5-x}$Sn$_x$ \cite{Kautzsch2023}, where the phase mode becomes soft.

The experimental setup we propose is shown in Fig.~\ref{fig:fig1}. In the iCDW phase, loop current fluctuations can be induced by an external laser pulse, exciting the phason modes. The overall geometry is similar to that used to probe magnon noise arising from non-collinear antiferromagnetic textures recently \cite{Finco2021}. Let us perform a qualitative estimation of the $T_1$ relaxation time induced by phason modes. According to Fermi's golden rule \cite{Rovny2024}:
\begin{eqnarray}
    T_1^{-1} &=& \frac{3}{2} \gamma_e^2 S_{B}(\omega_0),\\
    S_B(\omega) &=& \int dt e^{-i\omega t} \braket{B_{+}(t) B_{-}(0)}.
\end{eqnarray}
where $\omega_0$ is the frequency detected by the NV center, $B_{\pm} = B_{\hat{b}} \pm i B_{\hat{c}}$, and $\hat{b}, \hat{c}$ are the coordinate axes perpendicular to $\hat{a}$. Since the dynamics of the strip field $B(t)$ arise from the phase mode, we expect $B(t)$ to exhibit the frequency of $\omega_{ph}^{(A)}$. Therefore,
\begin{equation}
    T_{1}^{-1} \sim \frac{\gamma_e^2 |B|^2}{\omega_0 - \omega_{ph}^{(A)}}.
\end{equation}
Note that, according to the Biot-Savart law, $|B|$ depends on the distance of the NV tip from the sample. We take the local field \( B \) to be on the order of \( 0.01 \sim 0.1 \) mT, based on the orbital magnetization given in Ref.~\cite{Rina2024} (though it may differ in practice). Choosing \( \omega_0 - \omega_{ph} \sim 10 \) GHz, we estimate \( T_1 \sim 10 \text{ to } 1000 \,\mu\text{s} \), which falls within the sensitivity range of the NV center \cite{Finco2021, Rovny2024}.

\vspace{1\baselineskip}
\noindent{\bf Discussion}

In summary, we have provided a phenomenological analysis of the phase and amplitude modes of unconventional CDWs in Kagome lattice materials. Importantly, we have highlighted that the loop current order embedded in these unconventional CDWs can be probed with NV centers by exciting phase modes. We anticipate that the relevant experimental progress will significantly contribute to the search for loop current orders in real materials. Furthermore, it is also interesting to extend our experimental proposal to cuprates and explore the connection between loop current fluctuations and NV center detections in high-temperature superconductors \cite{Lee1991}.

 %\YM{possible application to high Tc}  	
\noindent {\bf Data availability} 

\noindent All data needed to evaluate the conclusions in the paper are present in the paper.

\noindent {\bf Acknowledgements}

\noindent  
N.N. was supported by JSPS KAKENHI Grant No. 24H00197 and 24H02231.
N.N. was also supported by the RIKEN TRIP initiative.  Y.M.X.  acknowledges financial support from the RIKEN Special Postdoctoral Researcher(SPDR) Program. %\YM{Hi nagaosa sensei, could you confirm the funding?}

\noindent {\bf Author contributions}

\noindent Y.M.X. and N.N. initiated this work. N.N. helped to analyze the problem. Y.M.X. carried out the calculations and wrote the manuscript with suggestions from N.N..

\noindent {\bf Competing interests}

\noindent The authors declare no competing interests.

  \clearpage
	
	\onecolumngrid
	\begin{center}
		\textbf{\large Supplementary Material for  ``Probing Loop Currents and Collective Modes of Charge Density Waves in Kagome Materials with NV Centers'' }\\[.2cm]
		Ying-Ming Xie,$^{1}$, Naoto Nagaosa,$^{1,2}$ \\[.1cm]
		{\itshape ${}^1$    RIKEN Center for Emergent Matter Science (CEMS), Wako, Saitama 351-0198, Japan}\\
        	{\itshape ${}^2$    Fundamental Quantum Science Program, TRIP Headquarters, RIKEN, Wako 351-0198, Japan}\\[1cm]
	\end{center}
	\setcounter{equation}{0}
	\setcounter{section}{0}
	\setcounter{figure}{0}
	\setcounter{table}{0}
	\setcounter{page}{1}
	\renewcommand{\theequation}{S\arabic{equation}}
	\renewcommand{\thetable}{S\arabic{table}}
	\renewcommand{\thesection}{\Roman{section}}
	\renewcommand{\thefigure}{S\arabic{figure}}
	\renewcommand{\bibnumfmt}[1]{[S#1]}
	\renewcommand{\citenumfont}[1]{#1}
	\makeatletter
	
	\onecolumngrid
	
	\maketitle

\section{Mean-field free energy from electron-electron interactions}
The ansatz order is given by

	\begin{eqnarray}
	\Delta_{CDW}(\bm{r})=\begin{pmatrix}
		0& \Delta_{\bm{Q}_3}(\bm{r})&   \Delta^*_{\bm{Q}_2}(\bm{r})\\
		 \Delta_{\bm{Q}_3}(\bm{r})^*&0&\Delta_{\bm{Q}_1}(\bm{r})\\
	\Delta_{\bm{Q}_2}(\bm{r})&\Delta^*_{\bm{Q}_1}(\bm{r})&0
	\end{pmatrix},\nonumber\\
\end{eqnarray}

where $\Delta_{\bm{Q}_j}(\bm{r})=|\Delta_{\bm{Q}_j}|e^{i(\bm{Q}_j\cdot \bm{r}+\theta_j)}$.  The basis is $(c_{A}, c_{B}, c_{C})$.

The patch Hamiltonian is
\begin{equation}
H(\bm{k})=\begin{pmatrix}
\epsilon_{1}(\bm{k})& |\Delta_{\bm{Q}_3}| e^{i\theta_3} &|\Delta_{\bm{Q}_2}| e^{-i\theta_2}\\
|\Delta_{\bm{Q}_3}| e^{-i\theta_3}& \epsilon_{2}(\bm{k})&|\Delta_{\bm{Q}_1}| e^{i\theta_1}\\
|\Delta_{\bm{Q}_2}| e^{i\theta_2}&|\Delta_{\bm{Q}_1}| e^{-i\theta_1}& \epsilon_{3}(\bm{k})
\end{pmatrix}
\end{equation}
where the dispersion near $M_{\alpha}$ points are $\epsilon_{1}(\bm{k})=-\mu-2\sqrt{3} tk_xk_y-2tk_x^2, \epsilon_{2}(\bm{k})=-\mu-3 tk_y^2+tk_x^2, ,\epsilon_{3}(\bm{k})=-\mu+2\sqrt{3} tk_xk_y-2tk_x^2 $. It is easy to verify that $\epsilon_3(C_3^{-1}\bm{k})=\epsilon_2(\bm{k})$, $\epsilon_2(C_3^{-1}\bm{k})=\epsilon_1(\bm{k})$, where $C_3$ is a three-fold rotation. The three-fold operator: \begin{equation}
  U_{C_3}=\begin{pmatrix}
      0&1&0\\
      0&0&1\\
      1&0&0
  \end{pmatrix}. 
\end{equation}
It is easy to show that  when $|\Delta_{\bm{Q}_1}|=|\Delta_{\bm{Q}_2}|=|\Delta_{\bm{Q}_3}|$ and $\theta_1=\theta_2=\theta_3\equiv \theta_0$, the Hamiltonian preserves  three-fold symmetry:
\begin{equation}
  U_{C_3} H(C_3^{-1}\bm{k}) U^{\dagger}_{C_3}= H(\bm{k}).
\end{equation}

Let us define the density operators
\begin{eqnarray}
\rho_{r}(\bm{k})&&=\sum_{\alpha\beta} \psi^{\dagger}_{\bm{k}_{\alpha}+\bm{k}/2}(\Gamma_r)_{\alpha\beta}\psi_{\bm{k}_{\beta}+ \bm{k}/2}, \\
\rho_{i}(\bm{k})&&=\sum_{\alpha\beta} \psi^{\dagger}_{\bm{k}_{\alpha}+\bm{k}/2}(\Gamma_i)_{\alpha\beta}\psi_{\bm{k}_{\beta}+ \bm{k}/2},
\end{eqnarray}
where $\psi_{\bm{k}}$ is the electron annihilation operator at momentum $\bm{k}$, and
\begin{equation}
\Gamma_{r}=\begin{pmatrix}
0&1&1\\
1&0&1\\
1&1&0
\end{pmatrix}, 
\Gamma_{i}=\begin{pmatrix}
0&1&-1\\
-1&0&1\\
1&-1&0
\end{pmatrix},
\end{equation}
and $\bm{k}_{\alpha}$ labels the momentum of $M_{\alpha}$. 
By introducing the Levi-Civita symbol $\epsilon_{\alpha\beta\gamma}$, the density operators can be rewritten as
\begin{eqnarray}
   \rho_{r,\alpha}(\bm{k})&&= \sum_{\beta\gamma}|\epsilon_{\alpha \beta \gamma}|\psi^{\dagger}_{\bm{k}_{\beta}+\bm{k}/2} \psi_{\bm{k}_{\gamma}+\bm{k}/2}\\
   \rho_{i,\alpha}(\bm{k})&&= \sum_{\beta\gamma}\epsilon_{\alpha \beta \gamma}\psi^{\dagger}_{\bm{k}_{\beta}+\bm{k}/2} \psi_{\bm{k}_{\gamma}+\bm{k}/2}
\end{eqnarray}

The density-density interaction that drives rCDW and iCDW is given by
\begin{eqnarray}
H_{rCDW}&&=-\frac{g_r}{2} \sum_{\alpha,\beta, \alpha',\beta'}  \psi^{\dagger}_{\bm{k}_{\alpha'}+\bm{k}/2}(\Gamma_r)_{\alpha'\beta'}\psi_{\bm{k}_{\beta'}+ \bm{k}/2} \psi^{\dagger}_{\bm{k}_{\alpha}+\bm{k}/2}(\Gamma_r)_{\alpha\beta}\psi_{\bm{k}_{\beta}+ \bm{k}/2} \delta_{k_{\alpha}+k_{\alpha'}, k_{\beta}+k_{\beta'}}\nonumber\\
&&=-\frac{g_r}{2} \sum_{\alpha}  \rho_{r,\alpha}(\bm{k}) \rho_{r,\alpha}(\bm{k}),\\
    H_{iCDW}&&=-\frac{g_i}{2} \sum_{\alpha}  \rho_{i,\alpha}(\bm{k}) \rho_{i,\alpha}(\bm{k}).
\end{eqnarray}
Here, $g_r$ and $g_i$ are the interaction for the rCDW and iCDW channel, respectively.
 The  Hubbard-Stratonovich transformation 
\begin{equation}
\exp(\int d\tau \frac{g_r}{2} \sum_{\alpha} \rho_{\alpha,r}  \rho_{\alpha, r})=\frac{1}{Z_0} \int \mathcal{D} \Delta_{r} \exp(\int d\tau  \sum_{\alpha} (-\frac{\Delta^2_{\alpha,r}}{2g_r}+\Delta_{\alpha,r} \rho_{\alpha, r}) ),
\end{equation}
where
\begin{equation}
Z_0=\int \mathcal{D}\Delta_{r}\exp(\int d\tau  \sum_{\alpha} (-\frac{\Delta^2_{\alpha,r}}{2g})),
\end{equation}
and $\Delta_{\alpha}\equiv \Delta_{\bm{Q}_\alpha}$. Then the partition function after the Hubbard-Stratonovich  transformation:
\begin{equation}
Z=\frac{1}{Z_0} \int \mathcal{D} \Delta_{r} \mathcal{D}\Delta_{i} \mathcal{D}\bar{\psi}\mathcal{D}\psi \exp[-\int d\tau \left(\bar{\psi}_{\alpha}( \partial_{\tau} \delta _{\alpha\beta}-(H_0)_{\alpha\beta}(\bm{k})) \psi_{\beta} + \sum_{l,\alpha} (\frac{\Delta^2_{\alpha,l}}{g_{l}}+\Delta_{\alpha, r} \rho_{\alpha, r})\right)]. 
\end{equation}
We can obtain the free energy
\begin{eqnarray}
F&&=\sum_{\alpha,l} \frac{1}{2g_l} |\Delta_{\alpha,l}|^2-\text{ln}(\text{Det}\hat{\mathcal{G}}^{-1}),\\
&&= \sum_{\alpha} \frac{1}{2g_r} |\Delta_{\alpha,r}|^2+ \sum_{\alpha} \frac{1}{2g_i} |\Delta_{\alpha,i}|^2-\text{tr}[\text{ln}\hat{\mathcal{G}}^{-1}]
\end{eqnarray}
The Green's function
\begin{eqnarray}
\hat{\mathcal{G}}^{-1}(i\omega,n)&&=-i\omega_n+H_0-\hat{\Delta}_{CDW}\\
&&=\begin{pmatrix}
-i\omega_n+\epsilon_1(
\bm{k})&-|\Delta_3| e^{i\theta_3}&-|\Delta_2| e^{-i\theta_2}\\
-|\Delta_3| e^{-i\theta_3}& -i\omega_n+\epsilon_2(\bm{k})&-|\Delta_1|e^{i\theta_1}\\
-|\Delta_2|e^{i\theta_2}&-|\Delta_1|e^{-i\theta_1}& -i\omega_n+\epsilon_3(
\bm{k})
\end{pmatrix}
\end{eqnarray}
In the vicinity of the phase transition, the order parameter $\Delta$ is small. We can use the trick,
\begin{eqnarray}
\text{tr}[\text{ln}\mathcal{G}^{-1})]&&=\text{tr}\text{ln}[\mathcal{G}_0^{-1}(1+ \mathcal{G}_0\hat{\Delta})]\\
&&\approx \text{tr}\text{ln}[\mathcal{G}_0^{-1}]-\frac{1}{2}\text{tr}[(\mathcal{G}_0\hat{\Delta})^2]+ \frac{1}{3}\text{tr}[(\mathcal{G}_0\hat{\Delta})^3]-\frac{1}{4} \text{tr}[(\mathcal{G}_0\hat{\Delta})^4].
\end{eqnarray}

Let us evaluate these term one by one:
\begin{equation}
\frac{1}{2}\text{tr}[(\mathcal{G}_0\hat{\Delta})^2]=\int_{\bm{k}}\sum_{n}(\frac{|\Delta_1|^2}{(i\omega_n-\epsilon_2(\bm{k}))(i\omega_n-\epsilon_3(\bm{k}))}+\frac{|\Delta_2|^2}{(i\omega_n-\epsilon_1(\bm{k}))(i\omega_n-\epsilon_3(\bm{k}))}+\frac{|\Delta_3|^2}{(i\omega_n-\epsilon_1(\bm{k}))(i\omega_n-\epsilon_2(\bm{k}))}).
\end{equation}
where $\omega_n=(2n+1) T$ is the Matsubara frequency, $T$ is the temperature. 
Due to the three-fold symmetry requirements,  the above coefficients for $|\Delta_{\alpha}|^2$ are the same:
\begin{eqnarray}
\lambda_1&&=\int_{\bm{k}}\sum_{n}(\frac{1}{(i\omega_n-\epsilon_2(\bm{k}))(i\omega_n-\epsilon_3(\bm{k}))}\nonumber\\
&&=\int_{\bm{k}} \frac{f(\epsilon_2(\bm{k}))-f(\epsilon_3(\bm{k}))}{\epsilon_{2}(\bm{k})-\epsilon_{3}(\bm{k})}.
\end{eqnarray}
The third order term is
\begin{equation}
\frac{1}{3}\text{tr}[(\mathcal{G}_0\hat{\Delta})^3]= \lambda_2|\Delta_1||\Delta_2||\Delta_3|\cos(\theta_1+\theta_2+\theta_3),
\end{equation}
where
\begin{eqnarray}
\lambda_2&&=-\sum_{n}\int_{\bm{k}} \frac{1}{(i\omega_n-\epsilon_1(\bm{k}))(i\omega_n-\epsilon_2(\bm{k}))(i\omega_n-\epsilon_3(\bm{k}))}\nonumber\\
&&=-\int_{\bm{k}} [\frac{f(\epsilon_1(\bm{k}))}{(\epsilon_1(\bm{k})-\epsilon_2(\bm{k}))(\epsilon_1(\bm{k})-\epsilon_2(\bm{k}))}+ \frac{f(\epsilon_2(\bm{k}))}{(\epsilon_2(\bm{k})-\epsilon_1(\bm{k}))(\epsilon_2(\bm{k})-\epsilon_3(\bm{k}))}+ \frac{f(\epsilon_3(\bm{k}))}{(\epsilon_3(\bm{k})-\epsilon_1(\bm{k}))(\epsilon_3(\bm{k})-\epsilon_2(\bm{k}))}].
\end{eqnarray}
The fourth order term is
\begin{equation}
\frac{1}{4} \text{tr}[(\mathcal{G}_0\hat{\Delta})^4]=\frac{1}{2} \int_{\bm{k}} \sum_{n} \frac{(\epsilon_1|\Delta_1|^2+\epsilon_2|\Delta_2|^2+\epsilon_3|\Delta_3|^2-i(|\Delta_1|^2+|\Delta_2|^2+|\Delta_3|^2)\omega_n)^2}{(i\omega_n-\epsilon_1)^2(i\omega_n-\epsilon_2)^2(i\omega_n-\epsilon_3)^2} \label{fourth}
\end{equation}

Then the free energy
\begin{eqnarray}
F
=-\lambda_1' \sum_{\alpha} |\Delta_{\alpha}|^2+b\sum_{\alpha} \cos 2\theta_{\alpha} |\Delta_{\alpha}|^2+ \lambda_2|\Delta_1||\Delta_2||\Delta_3|\cos(\theta_1+\theta_2+\theta_3)+u_1(\sum_{\alpha}|\Delta_{\alpha}|^4)+u_2\sum_{\alpha\neq  \beta} |\Delta_{\alpha}|^2 |\Delta_{\beta}|^2. \nonumber\\
\end{eqnarray}
where $\lambda_1'=\lambda_1-\frac{1}{4}(\frac{1}{g_r}+ \frac{1}{g_i})$, $b=\frac{1}{4}(\frac{1}{g_r}-\frac{1}{g_i})$, $u_1, u_2$ are determined according to Eq.~\eqref{fourth}, and we set $\Delta_{\alpha,r}= |\Delta_{\alpha}| \cos \theta_{\alpha}$, and  $\Delta_{\alpha,i}= |\Delta_{\alpha}| \sin \theta_{\alpha}$. The free energy is minimized when
\begin{equation}
    -\lambda_1'-|b|+\lambda_2\cos (3\theta_0)|\Delta_{Q}|+2(u_1+u_2)|\Delta_Q|^2=0 \label{Gap_equation}
\end{equation}

We assume the third-order term is much smaller than the second order. If $g_r> g_i$, the second order of free energy is minimized with $\theta_{\alpha}=0$  with $\lambda_2<0$ ($\theta_{\alpha}=\pi$  with $\lambda_2>0$).  The ground state is real CDW (rCDW) order. If $g_r<g_i$,   the second order of free energy is minimized with $\theta_{\alpha}=\frac{\pi}{2} \text{ or } -\frac{\pi}{2}$. The ground state is imaginary CDW (iCDW) order.
If $g_r= g_i$, the free energy is minimized through the cubic term but multiple phase $\theta$ would be degenerate at this order. In this case, both rCDW and iCDW can be mixed.

\section{Amplitude and phase modes}

The Lagrangian that includes the fluctuation part is
\begin{eqnarray}
\mathcal{L}&&=\kappa_0\sum_{\alpha}|\partial_{\tau}\Delta_{\bm{Q}_{\alpha}}|^2+ \kappa_1\sum_{\alpha}|\nabla \Delta_{\bm{Q}_{\alpha}}|^2+\sum_{\alpha} (b\cos 2\theta_{\alpha}-\lambda_1')|\Delta_{\bm{Q}_{\alpha}}|^2+\lambda_2|\Delta_{\bm{Q}_1}| |\Delta_{\bm{Q}_2}| |\Delta_{\bm{Q}_3}| \cos(\theta_1+\theta_2+\theta_3)+\nonumber\\
&&u_1(\sum_{\alpha}|\Delta_{\bm{Q}_\alpha}|^4)+u_2\sum_{\alpha\neq  \beta} |\Delta_{\bm{Q}_\alpha}|^2 |\Delta_{\bm{Q}_\beta}|^2.
\end{eqnarray}
Let us consider  the fluctuation
\begin{equation}
\Delta_{\bm{Q}_{\alpha}, \bm{q}}\approx \Delta_{\bm{Q}_\alpha}(1+\mathcal{A}_{\alpha}(\bm{q}))e^{i(\theta_0+\theta_{\alpha}(\bm{q}))}.
\end{equation}
The Lagrangian of the fluctuation part up to the quadratic (factor out $|\Delta_{\bm{Q}}|^2$)  is
\begin{eqnarray}
\mathcal{L}_{fluc}(\omega,\bm{q})&&=\sum_{\alpha}[(\kappa_1 \bm{q}^2-\kappa_0 \omega^2)( \mathcal{A}_{\alpha}(q) \mathcal{A}_{\alpha}(-q)+\theta_{\alpha}(q) \theta_{\alpha}(-q)]+2|b| \theta_{\alpha}(q) \theta_{\alpha}(-q)\nonumber\\
&&-(\lambda'_{1}+|b|) \mathcal{A}_{\alpha}(q) \mathcal{A}_{\alpha}(-q)+\lambda_2|\Delta_{Q}|\{\cos 3\theta_0 \sum_{\alpha>\beta} \mathcal{A}_{\alpha}(q) \mathcal{A}_{\beta}(-q)-\frac{1}{2}\cos (3\theta_0)((\sum_{\alpha}\theta_{\alpha}(q))(\sum_{\beta}\theta_{\beta}(-q))\nonumber\\
&&+\frac{1}{2}\sin (3\theta_0) [\sum_{\alpha} \mathcal{A}_{\alpha}(q) \sum_{\beta} \theta_{\beta}(-q)+  \sum_{\alpha}\mathcal{A}_{\alpha}(-q) \sum_{\beta} \theta_{\beta}(q) ]\}+(6u_1+2u_2) |\Delta_{Q}|^2 \sum_{\alpha} \mathcal{A}_{\alpha}(q)\mathcal{A}_{\alpha}(-q)\nonumber\\
&&+8u_2|\Delta_{Q}|^2\sum_{\alpha>\beta} \mathcal{A}_{\alpha}(q) \mathcal{A}_{\beta}(-q).
\end{eqnarray}
where $q=(\bm{q},\omega)$. The coupling term is only allowed when time-reversal symmetry is broken.

Let us make the linear combination of $A_{\alpha}(\bm{q})$ and $\theta_{\alpha}(\bm{q})$ with the  $C_3$ symmetry:
\begin{eqnarray}
\mathcal{A}^{(A)}_{q}&&=\frac{1}{\sqrt{3}}(\mathcal{A}_1(q)+ \mathcal{A}_2(q)+ \mathcal{A}_3(q)), \theta^{(A)}_{q}= \frac{1}{\sqrt{3}}(\mathcal{\theta}_1(q)+ \mathcal{\theta}_2(q)+ \mathcal{\theta}_3(q)),\\
\mathcal{A}^{(E_1)}_q&&=\frac{1}{\sqrt{2}}(\mathcal{A}_1(q)-\mathcal{A}_3(q)), \theta^{(E_1)}_q=\frac{1}{\sqrt{2}}(\theta_{1}(q)-\theta_3(q)),\\
\mathcal{A}^{(E_2)}_q&&=\frac{1}{\sqrt{6}}(\mathcal{A}_1(q)-2\mathcal{A}_2(q)+\mathcal{A}_3(q)), \theta^{(E_2)}_q=\frac{1}{\sqrt{6}}(\theta_{1}(q)-2\theta_2(q)+\theta_3(q)).
\end{eqnarray}

Then we can rewrite the  $\mathcal{L}_{fluc}(\omega,\bm{q})$ as
\begin{eqnarray}
\mathcal{L}_{fluc}(\omega,\bm{q})&&= (\kappa_1\bm{q}^2-\kappa_0\omega^2-(\lambda_1'+|b|)+(6u_1+2u_2)|\Delta_{Q}|^2+\lambda_2|\Delta_Q|\cos3\theta_0+8u_2|\Delta_{Q}|^2) \mathcal{A}^{(A)}_{q} \mathcal{A}^{(A)}_{-q}\nonumber\\
&&+(\kappa_1\bm{q}^2-\kappa_0\omega^2+2|b|-\frac{3}{2}\lambda_2|\Delta_{Q}|\cos(3\theta_0))\theta_{q}^{(A)} \theta_{-q}^{(A)}+\frac{3}{2}\lambda_2|\Delta_{Q}|\sin(3\theta_0) (\mathcal{A}^{(A)}_{q}\theta^{(A)}_{-q}+ \mathcal{A}^{(A)}_{-q}\theta^{(A)}_{q})+\nonumber\\
&&(\kappa_1\bm{q}^2-\kappa_0\omega^2-(\lambda_1'+|b|)+(6u_1+2u_2)|\Delta_{Q}|^2-\frac{1}{2}\lambda_2|\Delta_{Q}|\cos 3\theta_0- 4u_2|\Delta_{Q}|^2)(\mathcal{A}^{(E_1)}_{q}\mathcal{A}^{(E_1)}_{-q}+ \mathcal{A}^{(E_2)}_{q}\mathcal{A}^{(E_2)}_{-q})\nonumber\\
&&+(\kappa_1 \bm{q}^2-\kappa_0\omega^2+2|b|)(\theta_{q}^{(E_1)}\theta_{-q}^{(E_1)}+ \theta_{q}^{(E_2)}\theta_{-q}^{(E_2)}).
\end{eqnarray}
Here, 
 the relations $\sum_{\alpha}\mathcal{A}_{\alpha}(q)\mathcal{A}_{\alpha}(-q)= \mathcal{A}_{q}^{(A)}\mathcal{A}_{-q}^{(A)}+\mathcal{A}_{q}^{(E_1)}\mathcal{A}_{-q}^{(E_1)}+\mathcal{A}_{q}^{(E_2)}\mathcal{A}_{-q}^{(E_2)}$ and $\sum_{\alpha>\beta}{A}_{\alpha}(q)\mathcal{A}_{\beta}(-q)=\mathcal{A}_{q}^{(A)}\mathcal{A}_{-q}^{(A)}-\frac{1}{2}(\mathcal{A}_{q}^{(E_1)}\mathcal{A}_{-q}^{(E_1)}+\mathcal{A}_{q}^{(E_2)}\mathcal{A}_{-q}^{(E_2)})$ are used. 
 Using the gap equation Eq.~\eqref{Gap_equation}, we can rewrite 
\begin{eqnarray}
\mathcal{L}_{fluc}(\omega,\bm{q})&&= (\kappa_1\bm{q}^2-\kappa_0\omega^2+4(u_1+2u_2)|\Delta_Q|^2) \mathcal{A}^{(A)}_{q} \mathcal{A}^{(A)}_{-q}\nonumber\\
&&+(\kappa_1\bm{q}^2-\kappa_0\omega^2+2|b|-\frac{3}{2}\lambda_2|\Delta_{Q}|\cos(3\theta_0))\theta_{q}^{(A)} \theta_{-q}^{(A)}+\frac{3}{2}\lambda_2|\Delta_{Q}|\sin(3\theta_0) (\mathcal{A}^{(A)}_{q}\theta^{(A)}_{-q}+ \mathcal{A}^{(A)}_{-q}\theta^{(A)}_{q})+\nonumber\\
&&(\kappa_1\bm{q}^2-\kappa_0\omega^2-\frac{3}{2}\lambda_2\cos3\theta_0|\Delta_{Q}|+4(u_1-u_2)|\Delta_{Q}|^2)(\mathcal{A}^{(E_1)}_{q}\mathcal{A}^{(E_1)}_{-q}+ \mathcal{A}^{(E_2)}_{q}\mathcal{A}^{(E_2)}_{-q})\nonumber\\
&&+(\kappa_1 \bm{q}^2-\kappa_0\omega^2+2|b|)(\theta_{q}^{(E_1)}\theta_{-q}^{(E_1)}+ \theta_{q}^{(E_2)}\theta_{-q}^{(E_2)}).
\end{eqnarray}
If $\theta_0=0, \pi$, i.e., rCDW case,
\begin{eqnarray}
\omega^{(A)}_{\text{amp}}- \omega^{(E)}_{\text{amp}}&&= \frac{3}{2} \lambda_2 \cos3\theta_0 |\Delta_Q|+12u_2|\Delta_{Q}|^2\\
\omega^{(A)}_{\text{ph}}- \omega^{(E)}_{\text{ph}}&&=-\frac{3}{2}\lambda_2 \cos3\theta_0 
 |\Delta_{Q}|
\end{eqnarray}
If $\theta_0=\pm \frac{\pi}{2}$, i.e., iCDW case,
the eigenmodes of A-irreducible representation  is given by
\begin{equation}
\begin{vmatrix}
\kappa_1\bm{q}^2-\kappa_0(\omega^{(A)})^2+4(u_1+2u_2)|\Delta_Q|^2&\frac{3}{2}\lambda_2 \sin(3\theta_0)|\Delta_{Q}|\\
\frac{3}{2}\lambda_2 \sin(3\theta_0)|\Delta_{Q}|&  \kappa_1\bm{q}^2-\kappa_0(\omega^{(A)})^2+2|b|
\end{vmatrix}=0
\end{equation}
Then, mixed Higgs and phase mode energy are
\begin{eqnarray}
\kappa_0 (\omega_{\pm}^{(A)})^2=|b|+2(u_1+2u_2)|\Delta_Q|^2\pm \sqrt{(|b|-2(u_1+2u_2)|\Delta_Q|^2)^2+\frac{9}{4}\lambda_2^2|\Delta_Q|^2}.
\end{eqnarray}

\end{document}